\documentclass[fleqn,usenatbib]{mnras}

\usepackage[T1]{fontenc}
\usepackage{ae,aecompl}

\usepackage{graphicx}   
\usepackage{amsmath}    
\usepackage{amssymb}    

\title[Energy function and environment of FRBs]{Energy function, formation rate and low-metallicity environment of fast radio bursts}

\author[G. Q. Zhang \& F. Y. Wang]{
    G. Q. Zhang,$^{1}$
    F. Y. Wang,$^{1,2}$\thanks{E-mail: fayinwang@nju.edu.cn}
    \\
    $^{1}$School of Astronomy and Space Science, Nanjing University, Nanjing 210093, China\\
    $^{2}$Key Laboratory of Modern Astronomy and Astrophysics (Nanjing University), Ministry of Education, Nanjing 210093, China
}

\begin{document}
    \label{firstpage}
    \pagerange{\pageref{firstpage}--\pageref{lastpage}}
    \maketitle

    \begin{abstract}
        In this paper, we investigate the energy function, formation rate and
        environment of fast radio bursts (FRBs) using Parkes sample and
        Australian Square Kilometer Array Pathfinder (ASKAP) sample. For the
        first time, the metallicity effect on the formation rate is
        considered. If FRBs are produced by the mergers of compact binaries, the
        formation rate of FRBs should have a time delay relative to cosmic star
        formation rate (CSFR). We get the time delay is about 3-5 Gyr and the
        index of differential energy function $\gamma$ ($dN/dE\propto
        E^{-\gamma}$) is between 1.6 and 2.0 from redshift cumulative distribution. The value of $\gamma$ is similar to that
        of FRB 121102, which indicates single bursts may share the same physical
        mechanism with the repeaters.
        In another case, if the formation rate of FRB is proportional to the SFR without time delay,
        the index $\gamma$ is about 2.3. In both cases, we
        find that FRBs may prefer to occur in low-metallicity environment with $ 12 +
        \log(\rm{O/H}) \simeq 8.40$, which is similar to those of long
        gamma-ray bursts (GRBs) and hydrogen-poor superluminous supernovae
        (SLSNe-I).
    \end{abstract}

    \begin{keywords}
        radio continuum -- general -- stars: neutron
    \end{keywords}

    \section{Introduction}
    Fast radio bursts (FRBs) with anomalously high dispersion measure
    (DM) \citep{2007Sci...318..777L, 2013Sci...341...53T, 2015MNRAS.447..246P,
        2018PrPNP.103....1K, 2018arXiv181005836P} are mysterious radio
    transients, and have been detected at frequencies between
    400 MHz and 8 GHz by ground-based radio telescopes. By now, more than sixty FRBs
    have been discovered, only FRB 121102 and FRB 180814 are repeating
    \citep{2016Natur.531..202S,2016ApJ...833..177S,Amiri2019}. The localization of
    the FRB 121102 \citep{2017Natur.541...58C, 2017ApJ...834L...8M,
        2017ApJ...834L...7T} confirmed the cosmological origin of this
    source (at $z = 0.19$). If many redshifts of FRBs are measured by
    upcoming instruments, the combined redshift and DM can be used as
    cosmological purpose, including measuring the baryon number density
    \citep{2014ApJ...783L..35D, 2016Natur.530..453K}, measuring cosmic
    proper distance \citep{2017A&A...606A...3Y}, constraining the
    cosmological parameters \citep{2014PhRvD..89j7303Z,
        2014ApJ...788..189G, 2018ApJ...856...65W}, measuring the Hubble
    Constant and cosmic curvature if some repeating FRBs are
    gravitationally lensed \citep{2018NatCo...9.3833L}, probing compact
    dark matter through strong lensed FRBs \citep{2016PhRvL.117i1301M,
        2018A&A...614A..50W} and testing Einstein's Weak Equivalence
    Principle (WEP) \citep{2015PhRvL.115z1101W, 2018ApJ...860..173Y}.

    Because physical constraints on the progenitors of FRBs are few,
    there are many theoretical models. They generally fall into two categories:
    emission from compact binary merger \citep{2013ApJ...776L..39K,2013PASJ...65L..12T,2016ApJ...822L...7W,2016ApJ...827L..31Z,2016ApJ...826...82L,
        2018PASJ...70...39Y} and emission from a neutron star \citep{2010vaoa.conf..129P, 2014A&A...562A.137F,
        2014ApJ...780L..21Z, 2014ApJ...797...70K, Dai2016, 2016ApJ...826..226K, 2016MNRAS.457..232C,
        2017ApJ...841...14M}.
    Some progenitor models \citep{2014ApJ...780L..21Z, 2016MNRAS.458L..19C, 2016MNRAS.457..232C,
        2017ApJ...841...14M} connect FRBs
    and young neutron stars produced from supernovae or gamma-ray bursts
    (GRBs), so that their formation rate may track the cosmic star formation
        rate (CSFR). During the past years, some statistical studies
    of FRBs have been performed to constrain the models
    \citep{2016MNRAS.457.2530B, 2016MNRAS.458..708C,
        2016ApJ...826..226K, 2016MNRAS.461..984O, 2016MNRAS.461L.122L,
        2016ApJ...830...75V, 2017JCAP...03..023W, 2017ApJ...846L..27F, 2017AJ....154..117L, 2017RAA....17...14C,
        2018ApJ...858...89C, 2018MNRAS.474.1900M,Lu2019,Zhang2019,Wang2019}.

    The energy function, formation rate and burst environment are crucial constraints on the progenitor models.
    For example, the host galaxy of FRB 121102 is found to be a low-metallicity,
    star-forming dwarf galaxy \citep{2017ApJ...834L...7T}. More recently,
    FRB 171020 with the smallest recorded DM=114 pc cm$^{-3}$ discovered
    by Australian Square Kilometer Array Pathfinder (ASKAP) survey is
    possible associated with the Sc galaxy ESO 601-G036 \citep{2018ApJ...867L..10M}.
    This galaxy is also metal-poor, and shares similar properties with the galaxy hosting the
    repeating FRB 121102. However, whether the two galaxies represent the
    broader FRB population property is unknown.
    Meanwhile, observations show that both long GRBs and hydrogen-poor superluminous supernovae (SLSNe-I) exhibit a
    strong preference for low-mass, low-metallicity galaxies
    \citep{2006Natur.441..463F, 2014ApJS..213...15W, 2014ApJ...787..138L}. Here we
    provide a detailed study on the energy function, rate and environment of FRBs using the Parkes sample and ASKAP sample.

    This paper is organized as follows. In section 2, we present FRB
    samples and derive the pseudo redshifts for FRBs. In section 3,
    considering time delay and the metallicity of host galaxy, we fit
    the cumulative distribution with two FRB rate models. Finally,
    summary is given in section 4. In this paper, we adopt the
    $\Lambda$CDM model with $ H_0$ = 67.74   km s$ ^{-1} $ Mpc$ ^{-1} $,
    $ \Omega_b = 0.31 $ and $ \Omega_\Lambda = 0.69 $
    \citep{2016A&A...594A..13P}.

    \section{FRB samples}
    \label{sec:samples} We select the data from
    \url{http://www.frbcat.org}, which includes FRBs observed by Parkes,
    UTMOST, ASKAP and Arecibo, etc \citep{2016PASA...33...45P}. These
    telescopes have different central frequencies, bandwidths and
    thresholds. So it's unreasonable to put them together to investigate
    statistical nature of FRBs. We only consider FRBs observed by the
    same telescope. We select the FRBs observed by Parkes and ASKAP. The
    Parkes sample contains 28 FRBs, while the ASKAP sample has 23 FRBs.

    The DM of FRB can be divided into several parts
    \begin{equation}
    \label{eq:dm}
    \rm{DM} = \rm{DM}_{\rm{MW}} + \rm{DM}_{\rm{IGM}} + \frac{\rm{DM}_{\rm{host}} + \rm{DM}_{\rm{src}}}{1+z},
    \end{equation}
    where $\rm{DM}_{\rm{MW}}$, $\rm{DM}_{\rm{IGM}} $,
    $\rm{DM}_{\rm{host}}$ and $\rm{DM}_{\rm{src}}$ are the contributions
    from the Milky Way, the intergalactic medium (IGM), the host galaxy
    and the source. According the position of FRB, the
    $\rm{DM}_{\rm{MW}}$ can be inferred from the galactic distribution
    of free electrons. The values of $\rm{DM}_{\rm{MW}}$ have been given
    in the FRB catalog, which are based on the NE2001 model
    \citep{2002astro.ph..7156C}. The $\rm{DM}_{\rm{IGM}}$ can be
    calculated as
    \citep{2003PASJ...55..901I,2003ApJ...598L..79I,2014ApJ...783L..35D}
    \begin{equation}
    \label{eq:dmigm}
    {\rm{DM}}_{\rm{IGM}}(z) =  \frac{3cH_0\Omega_b}{8\pi G m_p} f_{\rm{IGM}} \int^z_0 \frac{H_0f_e(z')(1+z')}{H(z')}dz' ,
    \end{equation}
    where $f_{\rm{IGM}} \sim 0.83$ is the fraction of baryon mass in the
    IGM \citep{1998ApJ...503..518F} and $ f_e(z) $ is the number ratio
    between free electrons and baryons in IGM. $ f_e(z) $ can be
    calculated as
    \begin{equation}
    \label{eq:fe}
    f_e(z) \simeq \frac{3}{4} \chi_{e, H}(z) + \frac{1}{8} \chi_{e,He}(z),
    \end{equation}
    where $ \chi_{e, H}(z) $ and $ \chi_{e,He}(z) $ are the cosmic
    ionization of hydrogen and helium, respectively. Because the
    redshifts of FRBs are small, we can ignore the evolution of these
    parameters and take $ \chi_{e, H}(z) \simeq 1, \chi_{e,He}(z) \simeq
    1$. Finally, $ f_e(z)=7/8 $ is chosen in our calculation.

    However, the values of $\rm{DM}_{\rm{host}}$ and
    $\rm{DM}_{\rm{src}}$ are highly uncertain. The only verified host
    galaxy of FRB 121102 shows $55\, \rm{pc}\, \rm{cm}^{-3} \le
    \rm{DM}_{\rm{host}} + \rm{DM}_{\rm{src}} \le 225 $ pc
    cm$^{-3}$\citep{2017ApJ...834L...7T}. However, whether the
    properties of the host galaxies of repeating FRBs and non-repeating
    FRBs are similar is still in debate.
    In this paper, we adopt the distribution of DM given by
        \citet{2017ApJ...839L..25Y}. They use 21 FRBs to infer
        $\rm{DM}_{\rm{host}}+ \rm{DM}_{\rm{src}} =
        267.00^{+172.53}_{-110.68} \rm{pc}\, \rm{cm}^{-3}$. We simulate
        20000 points and use these points to obtain the pseudo redshifts
        through equations (\ref{eq:dm}) and (\ref{eq:dmigm}).
        It must be noted that there could be a correlation between the host DM and scattering, free-free absorption and other properties, which may also affect the pseudo redshifts.

    \section{Method and Results}
    \label{sec:method}

    \subsection{Cumulative redshift distribution}
\begin{figure}
        \centering
        \includegraphics[width=0.7\linewidth]{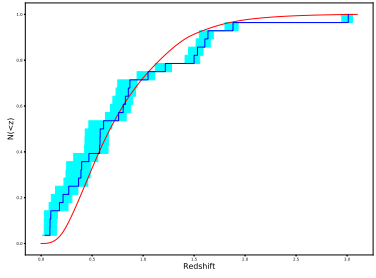}
        \caption{The cumulative distribution of 28 Parkes FRBs. The blue line is the observed distribution of FRBs and the shadow region
            is the uncertainties of the redshifts. The red line is the best-fitting result with $ \tau = 2.77 \rm{Gyr}, \gamma = 1.63 $ and $ Z = 0.46 Z_\odot $.}
        \label{fig:fig1}
    \end{figure}

\begin{figure}
        \centering
        \includegraphics[width=0.7\linewidth]{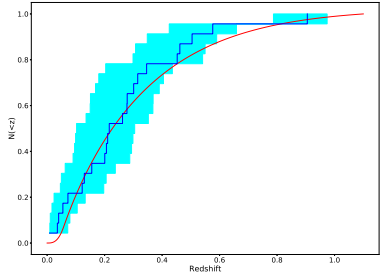}
        \caption{The cumulative distribution of 23 FRBs observed by ASKAP. The blue line is the distribution of FRBs and the shadow region
            is the uncertainties of the redshifts. Using the time delay $ \tau = 5.50 \rm{Gyr} $, $ \gamma = 2.07 $
            and the metallicity $ Z = 0.52 Z_\odot $, we give the best fitting as the red line.}
        \label{fig:fig2}
    \end{figure}

    Because the number of FRBs is small, it's preferable to use the
    cumulative distribution rather than differential distribution.
    It can avoid binning of the data.
    The width of binning has significant impact on the results when the
    number of events is not enough. According the pseudo redshifts given
    by equation (\ref{eq:dmigm}), we derive the cumulative redshift
    distributions for Parkes sample and ASKAP sample, which are shown in
    Figure \ref{fig:fig1} and Figure \ref{fig:fig2}, respectively. The
    blue line is the cumulative distribution and the shadow region is
    the uncertainties of the redshifts, which are caused by the
    uncertainty of $\rm{DM}_{\rm{host}} + \rm{DM}_{\rm{src}} $. It's
    obvious that the uncertainty of $\rm{DM}_{\rm{host}} +
    \rm{DM}_{\rm{src}} $ can slightly affect the cumulative
    distribution.

    We use two models to fit the cumulative redshift distributions. If
    FRBs origin from mergers of compact binaries, the formation rate
    should be proportional to the rate of compact binaries and has a
    time delay relative to CSFR. We use $ \dot{R}_m(z, \tau) $ as the
    rate of the compact binaries, where $ \tau $ is the time delay.
    Besides, according to the observation
    \citep{2017ApJ...834L...7T}, the metallicity of the
    host galaxy of FRB 121102 is poor. Thus we speculate that FRBs occur in
    low-metallicity environment. The metallicity evolution $ \Psi(Z, z)
    $ is considered. Therefore, the formation rate of FRB can be
    calculated as
    \begin{equation}\label{eq:rate}
    \rho_{FRB}(z) \propto \dot{R}_m(z, \tau)\Psi(Z, z).
    \end{equation}

    The second model is based on the progenitors model of FRBs
    associated with young neutron stars produced from supernovae or
    GRBs. In this case, the effect of time delay can be ignored and
    the formation rate is
    \begin{equation}
    \label{eq:4}
    \rho_{FRB}(z) \propto \rho(z) \Psi(Z, z),
    \end{equation}
    where $ \rho(z) $ is the CSFR.

    In order to derive the theoretical cumulative redshift distribution,
    the thresholds of telescopes must be considered. Assuming the energy
    distribution of FRBs satisfies the differential energy distribution
    $ \Phi(E)=dN/dE\propto E^{-\gamma} $, the theoretical cumulative
    redshift distribution can be obtained through
    \begin{equation}
    \label{eq:5}
    N(<z) = A \int_0^z \rho_{FRB}(z) [\int^{1}_{0} \eta(\varepsilon)\int_{E_{th} / \varepsilon}^{E_{max}} \Phi(E)dE d\varepsilon] \frac{dV(z)}{1+z},
    \end{equation}
    where $A$ is the normalized constant, $\varepsilon$ is the beam
        efficiency, $ \eta(\varepsilon)$ is the distribution of $\varepsilon$,
    $ E_{th} $ is the minimum
    energy which can be observed by telescopes and $ E_{max} $ is the
    maximum energy of FRBs. Below, we will discuss time delay,
    metallicity, beam efficiency and threshold in details.

    \subsection{CSFR and Time Delay}
    We adopt the CSFR given by \citet{2014ARA&A..52..415M}
    \begin{equation}
    \label{eq:6}
    \rho(z) = 0.015 \frac{(1 + z)^{2.7}}{1 + [(1 + z)/2.9]^{5.6}} \rm{M}_{\odot}~ \rm{year}^{-1}~ \rm{Mpc}^{-3} .
    \end{equation}
    If FRBs are produced by mergers of compact binaries, the formation
    rate should have a time delay to the CSFR. We take the time delay
        $\tau$ as $t(z) - t(z') $, where $z$ is the redshift
    when the compact binaries were formed, $z'$ is the redshift when the
    FRBs occurred and $t$ is the universe age at redshift $z$. The time
    delay $\tau$ should be determined by the time scale of the inspiral
    and formation process of compact binaries. There are many works
    to discuss the model of $\tau$. In our analysis, we take
    the probability distribution of $ \tau $ as $P(\tau)$ and an
    empirical expression of $ P(\tau) $ is taken as
    \citep{2018ApJ...858...89C}
    \begin{equation}
    \label{eq:7}
    P(\tau) \propto (\frac{\tau}{\tau_c})^{-1}e^{-\tau_c / \tau} ,
    \end{equation}
    where $\tau_c$ is a typical time scale. Considering the time delay,
    the rate of the compact binaries merger can be derived by
    \begin{equation}
    \label{eq:8}
    \dot{R}_m(z) \propto \int_0^{z_{max}} \rho(z') P(t(z) - t(z')) \frac{dt}{dz'} dz',
    \end{equation}
    where $dt / dz = -[(1 + z)H(z)]^{-1}$ and $ z_{max} $ is the maximum redshift
    when first stars formed.

    \subsection{Beam Efficiency}
        The positions of FRBs within the receiver beam of telescope
        have strong impact on the observed fluence. Therefore, for
        a multi-beam receiver, it is important to consider the beam efficiency.
        The Airy disk is adopted to illustrate the beam efficiency
        \citep{2016MNRAS.457.2530B,2016ApJ...830...75V},
        \begin{equation}
        \varepsilon = [\frac{2J_1(a)}{a}]^2,
        \end{equation}
        where $J_1(a)$ is the first order Bessel function and $a = r/r_c$
        represents the offset to the center of the main beam. The typical value of $
        \varepsilon $ is $ \varepsilon(a = 0) = 1$, and $\varepsilon(a =
        3.83) = 0 $.
        Considering the beam efficiency, the observed fluence
        is
        \begin{equation}
        \label{eq:beameff}
        F_{obs} = \varepsilon F_{eff},
        \end{equation}
        where $F_{eff}$ is the effective fluence which removes the beam effect.
        We also consider the probability
        distribution of $\varepsilon$ given by \citet{Niino2018ApJ...858....4N}
        \begin{equation}
        \label{eq:vardis}
        \eta(\varepsilon) = \frac{2}{a^2_{max}}a(\varepsilon)|\frac{da}{d\varepsilon}|,
        \end{equation}
        where $a_{max}$ is the maximum value of $a$ that satisfies
        $\varepsilon(a_{max}) = 0$, $a(\varepsilon)$ is the
        inverse function and $da/d\varepsilon$ is the differential
        function of $ a(\varepsilon) $.

        We adopt the Airy disk to describe the beam efficiency of Parkes.
        This is a simplification and the efficiency of radio receiver also
        depend on the frequency. Thus, the spectra of FRBs may affect the
        efficiency. In our analysis, a power-law spectrum is adopted
        for all FRBs. In each sample, all FRBs have the same spectrum,
        power-law index, central frequency and bandwidth. Therefore,
        the dependence can be ignored.

        As for the beam efficiency
        of ASKAP, due to its fly's-eye configuration, it's unacceptable
        to use a single airy disk to describe its beam efficiency.
        \citet{James2019PASA...36....9J} derived the best and worst beam efficiencies
        for closepack36 and square6$ \times $6 configurations. Because the contribution from closepack36 configuration is much larger
        than that of square6$ \times $6, we take the best beam efficiency from closepack36
        in our analysis.

    \subsection{Threshold}
    The thresholds of telescopes play an important role on the
    observed cumulative redshift distribution of FRBs. As for the
    Parkes, the fluence sensitivity is adopted as
    \citep{2016MNRAS.457.2530B}
    \begin{equation}
    \label{eq:9}
    F_{\nu, th} = 0.04 \frac{S/N}{\varepsilon} \sqrt{ \frac{\omega}{1 \rm{ms}}} \rm{Jy\, ms},
    \end{equation}
    where $S/N$ is the signal-to-noise, and we take it as 10. $\omega$
    is the typical duration of FRBs, which
    can be calculated from
    \begin{equation}
    \label{eq:10}
    \omega = \sqrt{\omega_{\rm{in}}^2(1 + z)^2 + \omega_{\rm{DM}}^2 + \omega_{\rm{sc}}^2},
    \end{equation}
    where $\omega_{\rm{in}}$ is the intrinsic width, $\omega_{\rm{DM}}$
    is the width caused by residual dispersion across a single frequency channel and $\omega_{\rm{sc}}$ is
    the contribution by the scattering. A typical intrinsic width
    $\omega_{\rm{in}} \simeq 1.3  $ ms is adopted. Assuming $ \Delta
    \nu_c / \nu_0 \ll 1 $, the $ \omega_{\rm{DM}} $ can be calculated as
    \citep{2016MNRAS.457.2530B}
    \begin{equation}
    \label{eq:11}
    \omega_{\rm{DM}} \simeq 8.3 \times 10^6 \frac{\rm{DM(z)} \Delta \nu_c}{\nu_0^3},
    \end{equation}
    where $\Delta \nu_c$ is the single frequency channel bandwidth and $\nu_0$ is the central
    frequency. As for the $\omega_{\rm{sc}}$, \citet{2013ApJ...776..125M}
    gave its form
    \begin{equation}
    \label{eq:12}
    \omega_{\rm{sc}} = \frac{k_{sc}[1 - \sqrt{z / (1 + z)}]}{\nu_0^4(1+z)} \times \int_0^z \frac{H_0}{H(z')}dz' \int_0^z \frac{H_0 (1+z')^3}{H(z')}dz'
    \end{equation}
    with $k_{\rm{sc}} = 4.2 \times 10^{13} $ ms MHz$^4$. According to eqs.
    (\ref{eq:9}) - (\ref{eq:12}), the threshold of Parkes can be
    obtained.

    For the ASKAP, we adopt 26 Jy ms as its threshold, which corresponds
    the signal-to-noise of 9.5 \citep{2018Natur.562..386S}. We don't
    consider the threshold evolution with redshift for ASKAP. It should
    be noted that the threshold of ASKAP is much larger than that of
    Parkes. So the redshifts of FRBs observed by ASKAP are small.
    Actually, The maximum pseudo redshift of FRBs observed by ASKAP is
    only 1.006. Thus we can ignore the threshold evolution.

    We assume that the differential energy distribution of FRBs
    satisfies a power-law form
    \begin{equation}
    \label{eq:13}
    \Phi(E) = \frac{dN}{dE} \propto E^{-\gamma}.
    \end{equation}
    In this equation, the energy $ E $ is not the intrinsic
    energy but the effective energy. Only the beam effect is considered
    in the effective energy, other selection effects such as propagation
    effect are not considered. These effects are complicated, and we
    have little understand of these. Considering the threshold of the
    telescopes, only FRBs with large energy can be observed. The minimum
    energy that can be observed at redshift $z$ is
    \begin{equation}
    \label{eq:14}
    E_{th} =  4 \pi d_c(z)^2 (1 + z) \Delta \nu F_{\nu,th}k(z),
    \end{equation}
    where $d_c$ is the comoving distance and $k(z)$ is the factor of the
    K-correction. The spectrum of FRB is still unclear. For simplicity,
    we assume that FRBs satisfy a power-law spectrum, $ F_\nu \propto \nu^{-\beta} $.
    Then $ k(z) $ is
    \begin{equation}\label{eq:15}
    k(z) = \frac{\nu_{\rm{max}}^{1 - \beta} - \nu_{\rm{min}}^{1 - \beta}}{[(1 + z)v_2]^{1 - \beta} - [(1 + z)v_1]^{1-\beta}},
    \end{equation}
    where $ \nu_{\rm{max}} $ and $ \nu_{\rm{min}} $ are the minimum frequency and maximum
    frequency at rest frame, respectively.
    As for Parkes, $ \nu_2 $ and $ \nu_1 $ are taken as 1522 MHz and 1182 MHz. The index of
    spectrum is uncertain. Previous studies have shown the index is $ \beta = 0.3 \pm 0.9 $
    for FRB 131104 \citep{2015ApJ...799L...5R} and $ \beta = 1.3 \pm 0.5 $ for FRB 150418
    \citep{2016Natur.530..453K}. Therefore, we take $ \beta = 0.8 $ for Parkes sample.
    The $\nu_2$ and $ \nu_1 $ is 1465 MHz and 1129 MHz for ASKAP. \citet{2018Natur.562..386S}
    obtained that the index of spectrum is $ 1.8 \pm 0.3 $ for ASKAP data. We also use
    this index for ASKAP. Assuming all FRBs redshifts are in the range of $0-4$, we obtain
    $ \nu_{\rm{max}} = 7610  $ MHz and $  \nu_{\rm{min}} = 1129 $ MHz.

    \subsection{Metallicity}
    Another important effect is the metallicity of host galaxy.
    \citet{2017ApJ...834L...7T} found that the host galaxy of FRB 121102
    is a low-metallicity galaxy. The 3$\sigma$ limit of its galaxy is
    $\rm{log}_{10}([O/H]) + 12 < 8.4$, which indicates that host
    galaxies of FRBs are more likely to be metal-poor galaxies.
    In order to
    describe the cosmic metallicity evolution, we employ the method developed
    by \citet{2006ApJ...638L..63L}. \citet{Panter2004MNRAS.355..764P} obtained the galaxy
    stellar mass function
    \begin{equation}
    \label{eq:massfunction}
    \psi(M) = A(\frac{M}{M_\odot})^\alpha e^{-M/M_\odot},
    \end{equation}
    where $\alpha = -1.16$, $A = 7.8 \times 10^{-3} h^3  \rm{Mpc}^{-3}$ and
    $M_\odot$ is the solar mass. Through this equation, the fraction of galaxies with
    mass less than $M$ can be calculated as
    \begin{equation}
    \label{eq:massfract}
    \Psi(M) = \frac{\int^M_0 M\psi(M)dM}{\int_0^\infty M\psi(M)dM} = \frac{\widehat{\Gamma}(\alpha + 2, M/M_\odot)}{\Gamma(\alpha +
    2)},
    \end{equation}
    where $\Gamma$ and $\widehat{\Gamma}$ is the complete and incomplete gamma
    function.
    Considering the mass-metallicity relation $ M/M_\odot =
        (Z/Z_\odot)^\zeta$ \citep{2005ApJ...635..260S} at $ z = 0.7 $
    and the evolution $Z = Z_\odot
    10^{-0.15z}$ \citep{2006ApJ...638L..63L}, the $\frac{M}{M_\odot}$ can be transformed into
    $\frac{Z}{Z_\odot}$ :
    \begin{equation}
    \label{eq:metaz}
    \Psi(Z, z) = \frac{\widehat{\Gamma}[\alpha + 2, (Z / Z_\odot)^\zeta 10^{0.15\zeta z}]}{\Gamma(\alpha + 2)},
    \end{equation}
    where  $\zeta = 2 $ and $ Z_\odot $ is the solar metallicity.
    This function describes
    the fraction mass density belonging to metallicities
    below metallicity $Z$ at redshift $z$ and has been used to investigate the CSFR and
    the gamma-ray bursts \citep{2008MNRAS.388.1487L,
        2009MNRAS.400L..10W,2012ApJ...749...68S,2013A&A...556A..90W}.
    The above equation is based
    on the galaxy mass function, mass-metallicity relation and metallicity evolution with redshift.
    These equations have some simplifications and the validity of these simplifications
    need to be examined. \citet{2008MNRAS.388.1487L} studied these simplifications in detail
    and found it's enough to adopt this analytical form of metallicity evolution \citep{Hao2013ApJ...772...42H}.
    Thus we use equation (\ref{eq:metaz}) and ignore its uncertainty.

    \subsection{Results}
\begin{figure}
        \centering
        \includegraphics[width=0.7\linewidth]{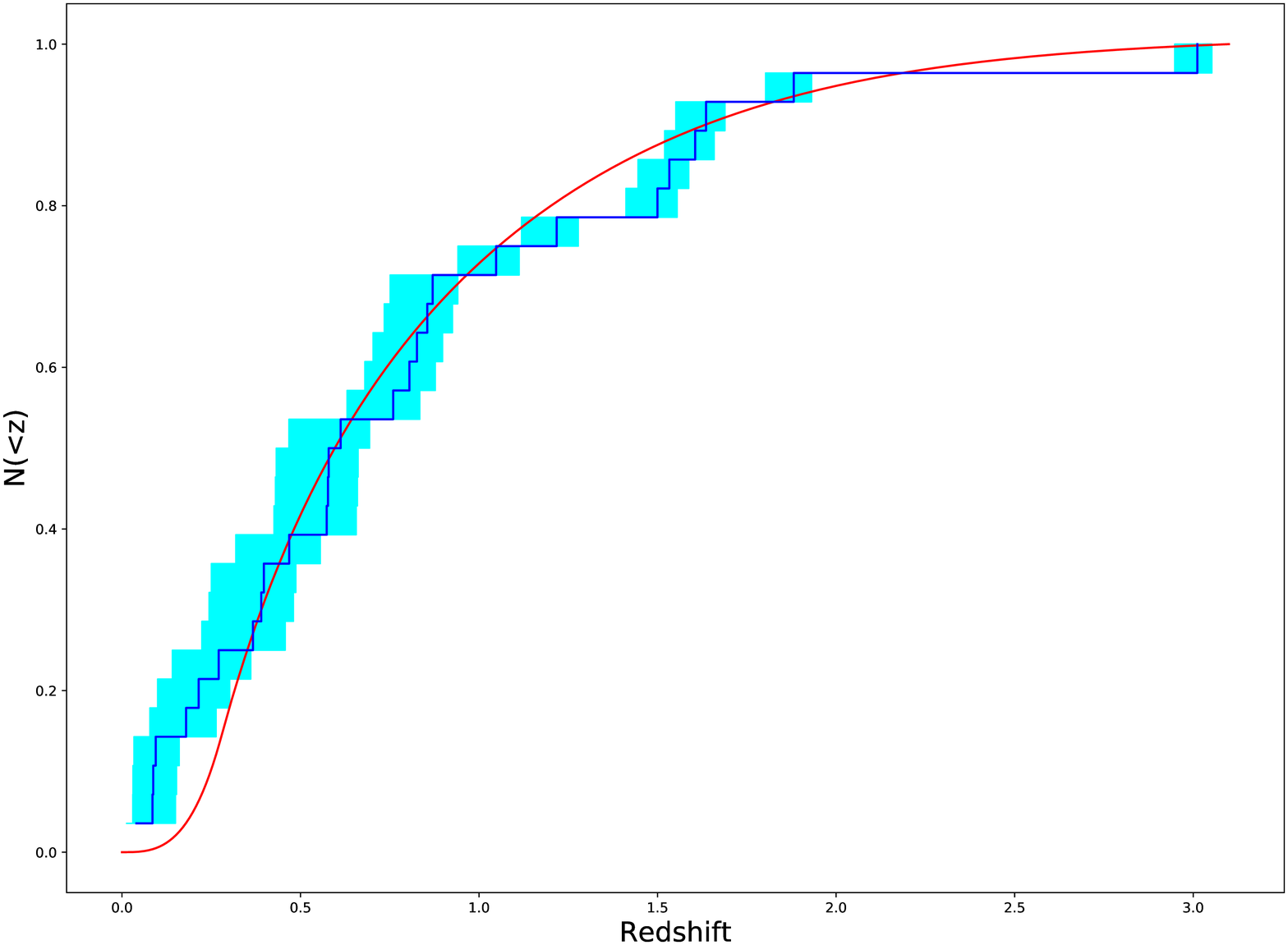}
        \caption{The cumulative distribution of 28 FRBs observed by Parkes. The blue line is the
            observed distribution of FRBs and the shadow regions is the uncertainties of the redshifts.
            The best-fitting is given by the red line with $ \gamma = 2.37 $ and $ Z = 0.52 Z_\odot $ }
        \label{fig:fig3}
    \end{figure}

\begin{figure}
        \centering
        \includegraphics[width=0.7\linewidth]{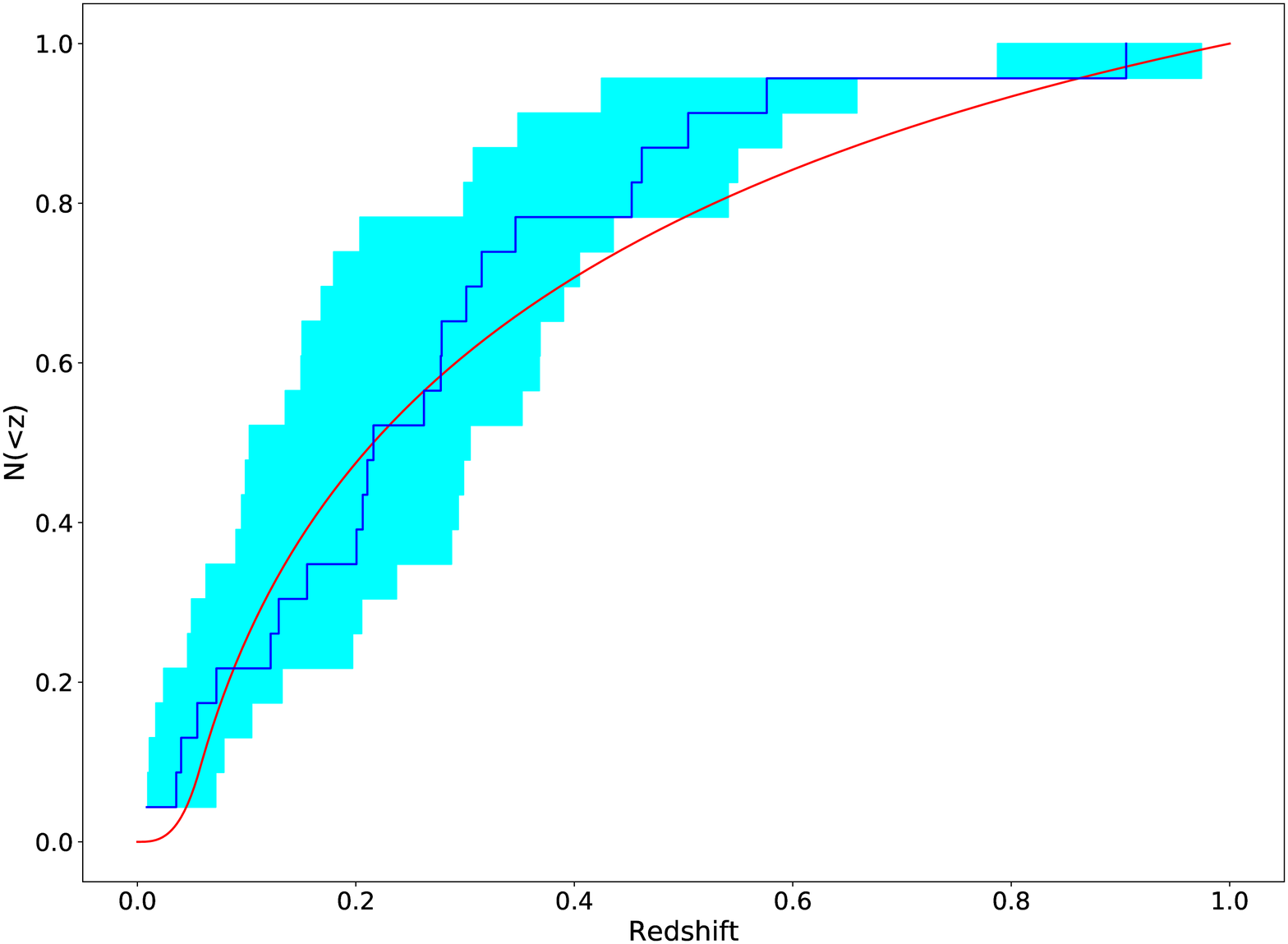}
        \caption{The cumulative distribution of 23 FRBs observed by ASKAP. The blue line is the observed distribution of FRBs and the shadow
            region is the uncertainties of the redshifts. The best-fitting is given by the red line with
            $ \gamma = 2.40$ and $ Z = 0.52 Z_\odot $.}
        \label{fig:fig4}
    \end{figure}

    We use Markov Chain Monte Carlo (MCMC) method to derive the best-fitting parameters.
    The time delay $ \tau $, the power-law index of energy function $ \gamma $ and
        the metallicity $ Z $ are taken as free parameters.
        According to the Bayes' theorem, the posterior probability can be derived through
    \begin{equation}
    \label{bayes}
        p(\theta|d) = \frac{p(d|\theta)p(\theta)}{p(d)},
    \end{equation}
    where $ \theta $ is the free parameters and $ d $ is the data of FRBs. We adopt
    the uniform distribution as the prior probability.
    In order to get the goodness of fit, we adopt the Kolmogorov-Smirnov
    test. The p-value of this test is considered as the likelihood.
    Through this likelihood, we derive the posterior probability for all
    cases. Considering the time delay, we obtain $ \tau =
        2.77^{+2.86}_{-1.90} $ Gyr, $ \gamma = 1.63^{+0.32}_{-0.25} $, $ Z =
        0.46^{+0.35}_{-0.31} Z_\odot $ for Parkes sample and $ \tau =
        5.50^{+3.01}_{-3.62} $ Gyr, $ \gamma = 2.07^{+0.14}_{-0.14} $, $ Z =
        0.52^{+0.32}_{-0.34} Z_\odot $ for ASKAP sample. The results are
    shown in figures \ref{fig:fig1} and \ref{fig:fig2}, respectively. In
    these figures, the blue line is observed distribution of FRBs and
    shadow region is the uncertainties of the redshifts. The red line is
    the best-fitting result. If the time delay is not considered, we
    find  $ \gamma = 2.37^{+0.12}_{-0.16} $, $ Z =
        0.52^{+0.34}_{-0.34} Z_\odot $ for Parkes sample and $ \gamma =
        2.40^{+0.08}_{-0.08}$, $ Z = 0.52^{+0.32}_{-0.34} Z_\odot $ for
        ASKAP sample.  The results are shown in figures \ref{fig:fig3} and
    \ref{fig:fig4}. We also list the best-fitting parameters in table
    \ref{tab:tab1}.

    The time delay derived from Parkes sample is $2.77^{+2.86}_{-1.90}$
    Gyr, which is similar to that for ASKAP sample
    $ \tau = 5.50^{+3.01}_{-3.62} $ Gyr at $ 1\sigma $ level. The large error of time delay is caused
    by the scarcity of the FRBs. \citet{2018ApJ...858...89C} also
    consider the time delay, but they obtained the time delay is about
    350 Myr, which is much smaller than ours. This inconsistency is due
    to the fact that they do not consider the metallicity effect.

    Another important result is the index of the differential energy
    distribution. Considering the time delay, the index for Parkes
    sample is $\gamma=1.63^{+0.32}_{-0.25}$, which is similar to the
    result of \citet{2018ApJ...858...89C}. Besides,
    \citet{2017JCAP...03..023W} and \citet{2017ApJ...850...76L} obtained
    that the index of the differential energy distribution for FRB
    121102 is also about 1.7. However, \citet{Gourdji2019} found the
    value of $\gamma=2.8\pm0.3$ for FRB 121102 using a sample of
    low-energy bursts. \citet{Wang2019} found a universal energy distribution
    with $\gamma \sim 1.7$ for FRB 121102 using bursts observed by different telescopes. As for ASKAP sample, the index is
    $2.07^{+0.14}_{-0.14}$. The difference between these two samples is
    caused by the selection effect. The threshold of ASKAP is much
    larger than that of Parkes, which causes the mean energy of ASKAP
    sample is larger. \citet{2016MNRAS.461L.122L} proposed that the
    index of differential energy distribution is $ 1.5 \le \gamma \le
    2.2 $ for the repeating FRB 121102. This value is consistent with
    our result. If the time delay is not considered, the indices are $
    2.37^{+0.12}_{-0.16} $ and  $ 2.40^{+0.08}_{-0.08} $ for Parkes and
    ASKAP samples, respectively.

    We find that the metallicity is low for all the cases. Through \
    $[12 + \log(\rm{O/H})]_{FRB} = [12 + \log(\rm{O/H})]_{sun} +
    \log(Z/Z_{\odot})$, the metallicity $[12 + \log(\rm{O/H})]$ are $ 8.35^{+0.25}_{-0.50} $,
    $ 8.40^{+0.21}_{-0.47} $, $ 8.41^{+0.22}_{-0.47} $, $
    8.38^{+0.24}_{-0.51} $ for Parkes sample with time delay, ASKAP
    sample with time delay, Parkes sample without time delay, ASKAP
    sample without time delay, respectively. We take $[12 +
    \log(\rm{O/H})]_{sun} = 8.69$ \citep{2009ARA&A..47..481A} as the
    solar metallicity. Although the time delay and energy distribution
    are different in the two models, the metallicities are similar.
    According to the observations, the host galaxies of FRB 121102
    \citep{2017ApJ...834L...7T} have low metallicity.
    Although whether the environment of repeating FRBs and
    non-repeating FRBs is the same is still unknown,
    our conclusion is consistent with the observation
    for repeating FRB 121102. Considering the large uncertainty, our results only provide a trend that the FRBs
    are more likely to occur in low-metallicity environment.
    The larger error of metallicity is mainly due to the small sample of FRBs. As more and more FRBs are observed, we expect
    more precise constraint on metallicity.

    Besides, observations confirmed that long GRBs and
    SLSNe-I prefer to occur in low-metallicity host galaxies
    \citep{2006Natur.441..463F,2014ApJ...787..138L}.
    The similar properties of host galaxy may indicate that they have
    similar progenitor model.

    \section{Summary And Discussion}
    \label{sec:conclusion}
    In this paper, we collect the FRBs observed
    by Parkes and ASKAP. We derive the pseudo redshifts from their DMs
    and obtain the cumulative distribution for each sample. Considering
    the effects of time delay and the metallicity, we construct the
    model of theoretical cumulative redshift distribution and obtain the
    best fit through MCMC method.

    We find the time delay is $2.77^{+2.86}_{-1.90}$ Gyr and $
    5.50^{+3.01}_{-3.62} $ Gyr for Parkes sample and ASKAP sample,
    respectively.  This time delay is very large and is consistent with
    the time delay for GRBs \citep{2015MNRAS.448.3026W}. The most
    important result is that FRBs prefer to occur in low-metallicity
    environment for all the cases for the first time. The cut-off
    metallicity is about $[12 + \log(\rm{O/H})]=8.30$. These results suggest that FRBs occur in
    low-metallicity environment, which is similar as those of GRBs and
    SLSNe-I.

    In our analysis, we use a simplified model to describe cosmic
    metallicity evolution and assume the metallicity is independent of
    the CSFR. However, there are some studies indicating that
    metallicity is connected with galaxy stellar mass and star formation
    rate in galaxy \citep{Mannucci2010MNRAS.408.2115M}. If this relation
    is rubust, the star formation rate has a strong effect on
    metallicity evolution, our assumptions may introduce bias. However,
    \citet{Sanchez2017MNRAS.469.2121S} proposed that this relation is
    not strong enough. Besides, this relation is only applied to the
    star formation rate of a galaxy. In our calculation, we adopt the
    CSFR rather than the star formation rate of a particular galaxy.
    Therefore, this relation is ignored in our analysis.

    As for the beam efficiency for ASKAP, we adopt the best efficiency for the
    closepack configuration given by \citet{James2019PASA...36....9J}. In their analysis,
    they provide 4 beam efficiencies, the best and worst beam efficiency for closepack36 configuration
    and square6$ \times $6 configuration. The square6$ \times $6 configuration is ignored, because its contribution is very small.
    We use the worst beam efficiency for the closepack configuration
    in ASKAP without time delay model to test whether beam efficiency
    can significantly affect our results. Using the worst beam efficiency, we obtain
    $ \gamma = 2.40^{+0.09}_{-0.08}, Z = 0.52^{+0.32}_{-0.34} $, which is consistent with the results for the best
    beam efficiency. This suggests the effect of different beam efficiencies can be ignored.

    It must be noted that we adopt a single power law to describe the spectrum of FRBs.
    This may import some uncertainties. In order to test the influence of spectrum, we also take the power-law indices
    $ \beta = 0,1.8$  for Parkes without time delay. We derive $ \gamma = 2.49^{+0.13}_{-0.13},
    Z = 0.53^{+0.32}_{-0.34} $ for $ \beta = 0 $
    and $ \gamma = 2.24^{+0.16}_{-0.18}, Z = 0.52^{+0.33}_{-0.35} $ for $ \beta = 1.8 $. The best-fitting
    results for different $ \beta $ are consistent with each other in $ 1\sigma $ confidence level.
    Therefore, the uncertainty of spectral indices doesn't significantly affect our results.
    Recently, some observations found that the intrinsic spectrum of FRBs may be not a single power law
    \citep{Hessels2019ApJ...876L..23H}. In this paper, we consider the spectra are power-law forms. If the spectra are not power laws, the above analysis should be
    reconsidered. In the future, the constraint on spectra will be more reliable. Our results can be tested with accurate spectrum.

    \section*{Acknowledgements}
    We thank the anonymous referee for useful suggestions
which were helpful for improving the manuscript. We also thank H. Qiu for helpful discussion on the beam efficiency for ASKAP. This work is supported by the National Natural Science Foundation of China
    (grant U1831207).


\clearpage
    \begin{table*}
        \centering
        \begin{tabular}{ccccc}
                            & Parkes (with time delay)   & ASKAP(with time delay) &  Parkes (without time delay) & ASKAP (without time delay)\\
            \hline
            $\tau$ (Gyr)    & $2.77^{+2.86}_{-1.90}$    & $5.50^{+3.01}_{-3.62}$    &                           &                           \\
            \hline
            $\gamma$        &  $1.63^{+0.32}_{-0.25}$   & $2.07^{+0.14}_{-0.14}$    & $ 2.37^{+0.12}_{-0.16} $  &  $  2.40^{+0.08}_{-0.08}$ \\
            \hline
            $Z (Z_\odot)$   & $0.46^{+0.35}_{-0.31}$    & $0.52^{+0.32}_{-0.34}$    & $ 0.52^{+0.34}_{-0.34} $  &  $  0.52^{+0.32}_{-0.34}$ \\
            \hline
            p-value         & 0.41                      & 0.92                      & 0.55                      & 0.78                      \\
            \hline
        \end{tabular}
        \caption{Best-fitting Results.}
        \label{tab:tab1}
    \end{table*}

    \bsp    
    \label{lastpage}
\end{document}